\documentclass[aps,pra,showpacs,10pt, twocolumn,tightlines,notitlepage,longbibliography]{revtex4-2}
\usepackage[colorlinks=true, citecolor=red, urlcolor=blue ]{hyperref}
\usepackage{times,amssymb,amsfonts,amsmath,bm,subfigure,mathtools,amsthm,braket,soul,enumitem,color, cancel, hyperref} 
\usepackage[utf8]{inputenc}
\usepackage{multirow}
\usepackage{graphicx}
\usepackage[dvipsnames]{xcolor}
\usepackage{tabularx}
\usepackage{comment}
\usepackage{pgfplotstable}
\usepackage{tikz}
\usetikzlibrary{quantikz2}
\usepackage{svg}
\usepackage{dsfont}
\title{AlgorithmTemplate}
\usepackage[linesnumbered,ruled,vlined]{algorithm2e}
\usepackage{fullpage}
\usepackage{fancyhdr}
\usepackage[ruled,vlined]{algorithm2e}

\pgfplotsset{width=10cm,compat=1.9}

\newcolumntype{C}[1]{>{\centering\arraybackslash}p{#1}}

\newcommand{\stkout}[1]{\ifmmode\text{\sout{\ensuremath{#1}}}\else\sout{#1}\fi}
\definecolor{magenta}{rgb}{1.0, 0.0, 0.56}

\newcolumntype{M}[1]{>{\centering\arraybackslash}m{#1}}
\newcolumntype{N}{@{}m{0pt}@{}}

\begin{document}
\title{Investigating layer-selective transfer learning of quantum approximate optimization algorithm parameters for the Max-Cut problem}
\author{Francesco Aldo Venturelli$^{1, 2, 3}$, Sreetama Das$^{4, 5, 6}$, Filippo Caruso$^{2,5,6}$}

\affiliation{$^1$Department of Engineering, University Pompeu Fabra, Barcelona, Spain}
\affiliation{$^2$Department of Physics and Astronomy, University of Florence, Via Sansone 1, Sesto Fiorentino, I-50019, Italy}
\affiliation{$^3$QUANTIC, Barcelona Supercomputing Center, Barcelona, Spain}
\affiliation{$^4$Institute for Cross-Disciplinary Physics and Complex Systems (IFISC) UIB-CSIC, Campus Universitat Illes Balears, Palma de Mallorca, Spain}
\affiliation{$^5$European Laboratory for Non-Linear Spectroscopy (LENS), University of Florence, Via Nello Carrara 1, Sesto Fiorentino, I-50019, Italy}
\affiliation{$^6$Istituto Nazionale di Ottica del Consiglio Nazionale delle Ricerche (CNR-INO), I-50019 Sesto Fiorentino, Italy}
\begin{abstract}
The quantum approximate optimization algorithm (QAOA) is a variational quantum algorithm (VQA) ideal for noisy intermediate-scale quantum (NISQ) processors, and is highly successful in solving combinatorial optimization problems (COPs). It has been observed that the optimal parameters obtained from one instance of a COP can be transferred to another instance, resulting in generally good solutions for the latter. In this work, we propose a refinement scheme in which only a subset of QAOA layers is optimized following parameter transfer, with a focus on the Max-Cut problem. Our motivation is to reduce the complexity of the loss landscape when optimizing all the layers of high-depth QAOA circuits, as well as to reduce the optimization time. We investigate the potential hierarchical roles of different layers and analyze how the approximation ratio scales with increasing problem size. Our findings indicate that the selective layer optimization scheme offers a favorable trade-off between solution quality and computational time, and can be more beneficial than full optimization at a lower optimization time.
\end{abstract}

\maketitle

\section{Introduction}
Combinatorial optimization problems (COPs) represent a large class of optimization problems relevant in many real-world scenarios, some examples being the traveling salesman problem, minimum vertex covering, graph coloring, etc. Due to the NP hardness of these problems, finding an optimal solution is challenging for a classical exact or heuristic algorithm. Specifically, the complexity of searching for the optimal solution(s) among feasible ones could exponentially increase as the problem scale grows~\cite{zhang2023review}. Quantum computation, which utilizes inherent quantum properties e.g. superposition and entanglement, has shown theoretical evidence of quantum advantages in specific class of problems~\cite{shor, grover1996, teleport_1993, BB84}.
Thus, significant interest has been focused on the development of specific algorithms, aimed at being executed on digital quantum computers,
with the final goal of surpassing their classical counterparts. The quantum approximate optimization algorithm (QAOA), suitable for solving COPs, was first proposed in Ref.~\cite{farhi2014qaoa}. QAOA was inspired by the quantum adiabatic theorem, which 
is also the underlying principle in quantum annealing (QA)~\cite{rajak_2023}, and QAOA can be perceived as Trotterized QA.  It is well known that the solution of a COP can be encoded in the ground state of a classical cost Hamiltonian $H_{c}$~\cite{lucas_2014} representing the instance of the problem itself. QAOA works by variationally preparing an approximate ground state of $H_{c}$.
It has been established that for certain classes of COPs, QAOA provides some quantum advantages over classical algorithms~\cite{farhi2014qaoa, farhi2019quantumsupremacyquantumapproximate, Hadfield_2019, Lykov2023, shaydulin_2024, BLEKOS20241}.

The QAOA algorithm belongs to the large class of circuit-based quantum machine learning algorithms, known as variational quantum algorithms (VQAs)~\cite{peruzzoVQA2014, McClean_2016},
which are ideal to implement in noisy intermediate-scale quantum (NISQ) devices. A subclass of these algorithms are variational quantum eigensolvers (VQEs), which are used to variationally prepare the ground state of a Hamiltonian $H_{c}$. In VQE, the gate parameters $(\vec{\theta}=\{\theta_{1}, \theta_{2},.., \theta_{m}\})$ of a quantum circuit $U(\vec{\theta})$ are iteratively updated using a classical optimization routine until the state $|\psi\rangle$ prepared by $U(\vec{\theta})$ minimizes the expectation value $\langle \psi\vert H_{c} \vert \psi\rangle$ of $H_{c}.$ 
In particular, for QAOA, $U(\vec{\theta})$ is composed of two unitary operators generated from $H_{c}$ and another ``mixer" Hamiltonian $H_{m}$, applied in an alternative order. A $p$-layer or $p$-depth QAOA has in total $2p$ such unitary applications and 
trainable parameters. We discuss the architecture in more detail in the next section. The approximate solution obtained from a finite-depth QAOA approaches the true solution as $p\rightarrow \infty$. With an increasing number of qubits, one requires a higher number of layers $p$ to obtain a good approximate solution. However, training a large-depth variational circuit can suffer from local minima~\cite{anschuetz_2022} and is prone to the barren plateau phenomena~\cite{mcclean_BP_2018} where the variance of the cost function gradient approaches zero exponentially in the number of qubits. To avoid these bottlenecks, several works have emphasized the importance of choosing certain underlying structures of parameters, as well as efficient parameter initialization~\cite{zhou_2020, Streif_2020, Sack2021quantumannealing, Lee_2021}. 

An efficient technique to bypass the training issue of QAOA circuit is to take advantage of the good transferability of QAOA parameters, which is based on the observation that the optimum parameters for one instance of a particular COP result in good quality solutions for other instances of that COP~\cite{brandao2018, galda_2021, galda2023similarity, shaydulin_2023, sakai2024linearly, lyngfelt_2025}. 
Furthermore, parameter transferability has been observed to be consistent even between different COPs~\cite{montanezbarrera2024}. Thus, the optimized QAOA parameters can be transferred from a particular instance with a smaller problem size to QAOA circuits corresponding to newer target instances with higher problem sizes, overcoming the resource-intensive optimization process for the latter. Despite this advantage, a detailed look at previous works suggests that the transferability of parameters is reduced with a growing difference in size and in other properties between the optimized instance and the target instance~\cite{akshay_2021, montanezbarrera2024, sakai2024linearly}. Therefore, another practical approach is to use the transferred parameters as initial parameters, followed by optimizing all of them to get a reasonably good solution. This approach, known as the `warm start' of QAOA parameters, reduces the optimization time compared to the self-optimization in which the initial parameters are chosen randomly.

In this work, we propose a more generalized QAOA transfer learning scheme for Max-Cut problem. In this scheme, after the parameter transfer, we optimize only a small subset of layers to further improve the solution quality. The motivation of our approach is to significantly reduce the optimization time compared to optimizing all layers, as well as to investigate whether there exists some bias in the optimization landscape due to which selectively optimizing some of the layers can lead to better performance than optimizing others. The performance obtained from this approach is evidently lower than that obtained from all-layer optimization, with the benefit of a shorter optimization time. Additionally, we study the scaling behavior of the selective optimization with increasing size of the Max-Cut graphs. Our findings show that for a majority of Max-Cut instances, there exists a hierarchy among the layers when selectively optimizing them.

The manuscript is organized as follows. In Sec.~\ref{prelim}, we describe the QAOA algorithm, the Max-cut problem, and the concept of parameter transfer in QAOA circuit. In Sec.~\ref{results} we present our numerical results, and in Sec.~\ref{conclusion} we present the conclusions.

\begin{figure*}[t!]
    \centering
    \includegraphics[width=0.95\textwidth]{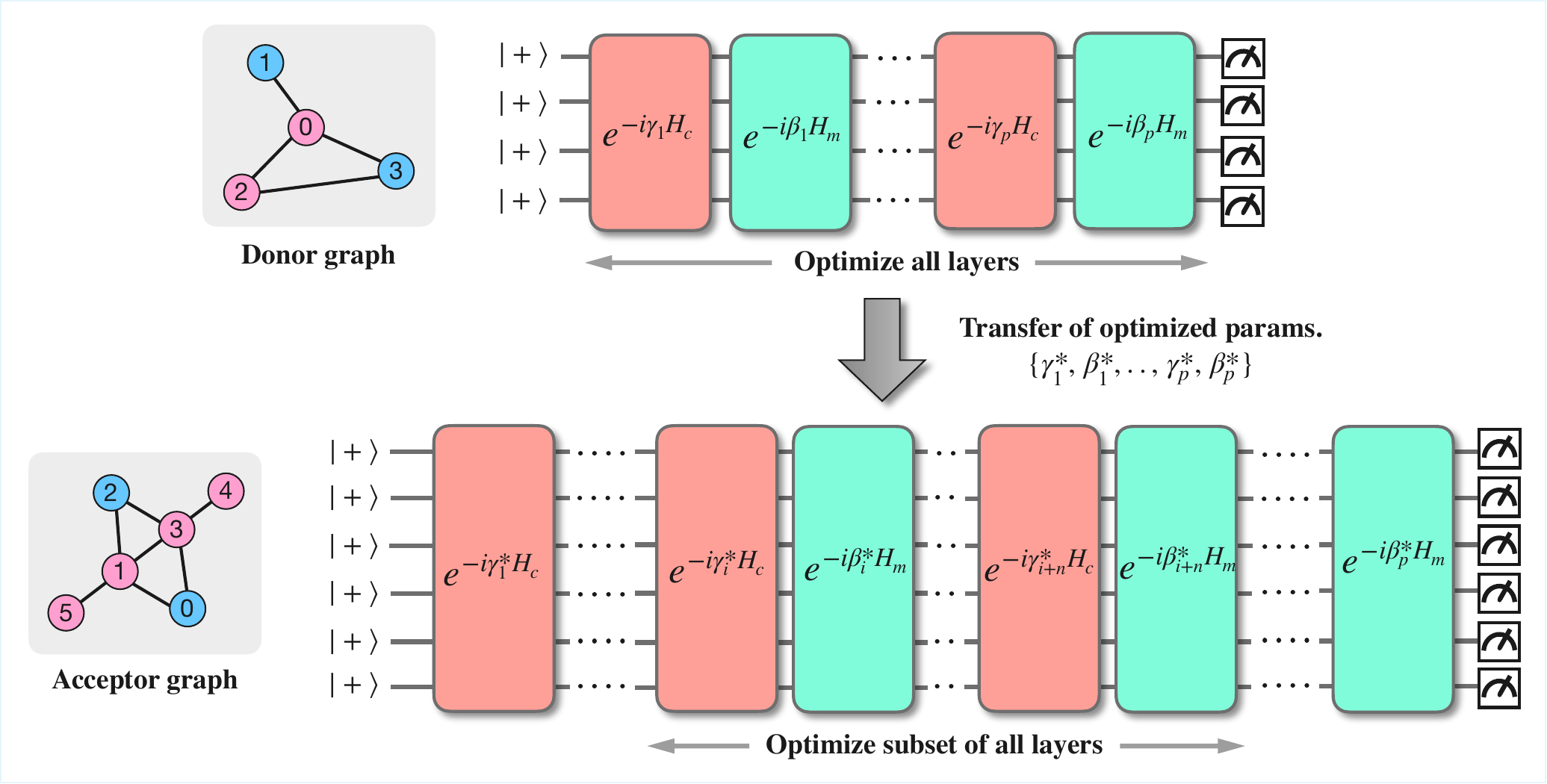}
    \caption{A schematic diagram of the layer-selective transfer learning scheme of QAOA for the Max-Cut problem. We show a donor and an acceptor graph, where the blue and pink colors of the nodes denote the binary labeling corresponding to the Max-cut solution. The parameters of a $p$-layer QAOA circuit are optimized for the donor graph. The optimized parameters $\{\gamma_{k}^{*}, \beta_{k}^{*}\}$ ($k = 0, 1, 2, .., p$) are transferred to the QAOA circuit of the acceptor graph, following which only $n$ ($n<p$) layers with initial parameters $\{\gamma_{i}^{*}, \beta_{i}^{*}, ..., \gamma_{i+n}^{*}, \beta_{i+n}^{*}\}$ are updated to minimize the cost function for the acceptor graph. For presentation purpose, we show the selective training of the intermediate $n$ layers; however, the scheme can be generalized to any set of $n$ layers.}
    \label{qaoa_schematic}
\end{figure*}

\section{Methods}
\label{prelim}
\subsection{QAOA algorithm}
QAOA can be interpreted as a Trotterized QA algorithm, assisted by a classical optimization routine. Similar to QA, the time evolution in QAOA is generated by two Hamiltonians, namely the cost Hamiltonian $H_{c}$ which encodes the specific COP to solve, and the mixer Hamiltonian $H_{m}$. The cost Hamiltonian is chosen in a way so that its ground state encodes the solution to the COP. Usually, $H_{c}$ is diagonal in the computational basis, and one (or more) of these basis states corresponds to the solution. On the other hand, $H_{m}$ is non-diagonal in the computational basis and $[H_c, H_m]\neq 0$. 
The initial state of QAOA is an eigenstate $\vert \Psi_{0}\rangle$ of $H_{m}$. The most commonly used mixer Hamiltonian is
\begin{equation}
    H_{m} = \sum\limits_{i=0}^{N-1}\sigma^{X},
\end{equation}
with the corresponding initial state $\vert \Psi_{0}\rangle = |+\rangle^{\otimes N}$ for $N$ qubits. Here, $N$ is the number of qubits that is equivalent to the size of the COP, $\sigma^{X}$ is the Pauli-X operator and $\vert +\rangle = \frac{1}{\sqrt{2}}(|0\rangle + |1\rangle)$. Depending on the particular COP and the knowledge of a feasible solution set, it may be prudent to use other mixer Hamiltonians~\cite{Hadfield_2019, wang_2020, cook_2020, eidenbenz_grover_2020}.
To construct the variational ansatz, one alternately applies the unitaries 
\begin{equation}
    U(H_c, \gamma_i) = e^{-\mathrm{i}\gamma_{i} H_{c}}, \;\; U(H_m, \beta_i) = e^{-\mathrm{i}\beta_{i} H_{m}}
\end{equation}
on $\vert \Psi_{0}\rangle$. Here $\{\gamma_{i}, \beta_{i}\}$ are the durations for which these two unitaries act in the $i$th layer. Therefore, optimizing these parameters implies optimizing the time scheme of evolution under the above two unitaries. Together, they constitute the $2p$ trainable parameters in a $p$-layer QAOA. Application of $U(H_c, \gamma_i)$ introduces phases between different basis states of $\vert \Psi_0\rangle$, whereas application of $U(H_m, \beta_i)$ creates interference between these phases. When applied alternatively several times with the proper set of values of $\{\gamma_{i}, \beta_{i}\}$, these unitaries are responsible for the evolution of the state $|\Psi_{0}\rangle$ toward the desired solution. After applying $p$ such layers, the final state is
\begin{align}
    |\Psi\rangle &= U(H_m,\beta_p)U(H_c,\gamma_p)\;\dots \\ \nonumber 
    &U(H_m, \beta_2)U(H_c, \gamma_2)U(H_m, \beta_1)U(H_c, \gamma_1)|\Psi_0\rangle.
\end{align}

After each circuit simulation, one measures the expectation value $\langle H_{c}\rangle = \langle \Psi | H_{c} |\Psi\rangle$, and it is minimized by repetitively updating $\gamma$'s and $\beta$'s using a classical optimization routine, which can be gradient-based or gradient-free. The performance of QAOA is evaluated using the approximation ratio $r$ defined as
\begin{equation}
    r = \frac{\langle H_{c}\rangle_{\mathrm{min}}}{ E_{\mathrm{min}}},
\end{equation}
where $E_{\mathrm{min}}$ is the true ground state energy, and $\langle H_{c}\rangle_{\mathrm{min}}$ is the minimum cost function obtained from the classical optimization. Upon completion of the optimization process, $\vert \Psi\rangle$ is a superposition of computational basis states, from which one can sample the solution state with high probability by projectively measuring all the qubits. For certain classes of COPs, the QAOA circuit with depth $p$\;=\;$1$ is guaranteed to find the solution with $\sim 69\%$ approximation ratio~\cite{farhi2014qaoa}. The output state $\vert \Psi\rangle$ approaches the true solution as $p \rightarrow \infty$, in which limit $r=1$. In general, it is not clear exactly how many layers one must optimize to guarantee a desired value of $r$.

\subsection{Max-Cut problem}
The maximum cut problem or Max-Cut refers to labeling the nodes of a graph into two groups in a way to maximize the total number of graph edges connecting two nodes from different groups. Solving Max-Cut for unweighted graphs is equivalent to finding the ground state of the Hamiltonian
\begin{equation} 
    H_{c} = \frac{1}{2}\;\sum_{\mathclap{\langle i, j\rangle \in S}}\big(\mathds{1} - \sigma^{Z}_{i}\sigma^{Z}_{j}\big)\;,
\end{equation}
where $\langle i, j\rangle$ denotes an edge connecting the $i$th and $j$th node, and $S$ is the set of all edges.
For weighted graphs, the Max-Cut problem can be mapped to the following Hamiltonian
\begin{equation} 
    H_{c} = \frac{1}{2}\sum_{\langle i, j\rangle \in S}J_{ij}\big(\mathds{1}- \sigma^{Z}_{i}\sigma^{Z}_{j}\big).
\end{equation}
Here, $\sigma^{Z}_{i}$ is the Pauli-Z matrix acting on $i$th qubit, and $J_{ij}>0$ are edge weights randomly generated from a uniform distribution.

\subsection{Transferability of QAOA parameters}
In Ref.~\cite{brandao2018}, a concentration behavior of the QAOA cost function corresponding to the Max-Cut problem was heuristically and numerically observed for 3-regular graphs. In detail, for a fixed depth $p$ of the QAOA circuit and in the limit of a large number of qubits $N$, the cost function landscape and the optimum parameter values have very little dependence on the particular chosen instance of the Max-Cut problem. Thus, the optimal parameter values for different cases concentrate in the same region of the parameter space. For non-isomorphic and unweighted graphs, similar observations were reported in Ref.~\cite{Lotshaw2021}. The concentration phenomena and the scaling of the parameters were analytically demonstrated in Ref.~\cite{akshay_2021}. These observations lead to the scheme of `parameter transfer', that is, optimizing QAOA parameters for one instance (donor graph) and then transferring them to the QAOA circuit for a different instance (acceptor graph), without optimizing them further for the latter. Many other works studied in detail the effectiveness of the optimum parameter transfer for smaller problem sizes, between graphs with varying topological properties, for the weighted Max-Cut problem, and even between instances of different COPs~\cite{galda_2021, galda2023similarity, shaydulin_2023, sakai2024linearly, montanezbarrera2024}.

\begin{algorithm}[]
\caption{General procedure to optimize a subset of $n$ layers ($n \leq p$) in a $p$-layer QAOA\\circuit after transferring parameters from an $N_1$-node donor graph $G_1$ to an $N_2$-node acceptor graph $G_2$ ($N_2 \geq N_1$). This algorithm generalizes the single-layer optimization scheme.}
\label{our_algo}
\SetAlgoLined
\vspace{0.2cm}
\KwData{\\
  Graphs $G_{1} \in \mathcal{G}(N_{1}, E_{1})$,
  $G_{2} \in \mathcal{G}(N_{2}, E_{2})$, 
  number of layers $n$ to optimize, maximum number of iteration steps $T$, tolerance tol$=10^{-4}$, integer $s$ for detecting convergence.
}
\vspace{0.1cm}
\KwResult{\\
  Optimized QAOA parameters $\theta_{G_2}^p$ for $G_2$:
  \[
  \theta_{G_2}^{p} = \{(\gamma_{1}, \beta_{1}), \dots, (\gamma_{n}^{*}, \beta_{n}^{*}), \dots, (\gamma_p, \beta_p)\}
  \]
}
\vspace{0.1cm}
Solve Max-Cut on $G_1$ to obtain $\theta_{G_1}^{p} = \{(\gamma_1, \beta_1), \dots, (\gamma_p, \beta_p)\}$ \\
Initialize $\theta_{G_2} \leftarrow \theta_{G_1}$ \\
Freeze all layers $\theta_{G_2}^i$ except for the $n$th \\
Compute initial cost $H_0 \leftarrow H_{t=0}(\theta_{G_2})$ \\
$early\_stop \leftarrow 0$ \\
\For{$t \leftarrow 1$ \KwTo $T$}{
    Compute gradient: $\nabla H_t \leftarrow \nabla_{\theta^{n}} H_t(\theta_{G_2})$ \\
    Update parameters $\theta_{G_2}$ using gradient step \\
    Compute cost $H_t \leftarrow H(\theta_{G_2})$ \\
    \If{$|H_t - H_{t-1}| < \mathrm{tol}$}{
        $early\_stop \leftarrow early\_stop + 1$ \\
        \If{$early\_stop > s$}{
            \textbf{break}
        }
    }
    \Else{
        $early\_stop \leftarrow 0$
    }
    $H_{t-1} \leftarrow H_t$ \\
}
\end{algorithm}

\begin{figure*}[t]
\hspace{-1cm}
\centering
\begin{tikzpicture}

    \node[anchor=south west,inner sep=0] (image) at (0,0) {\includegraphics[width=0.8\textwidth]{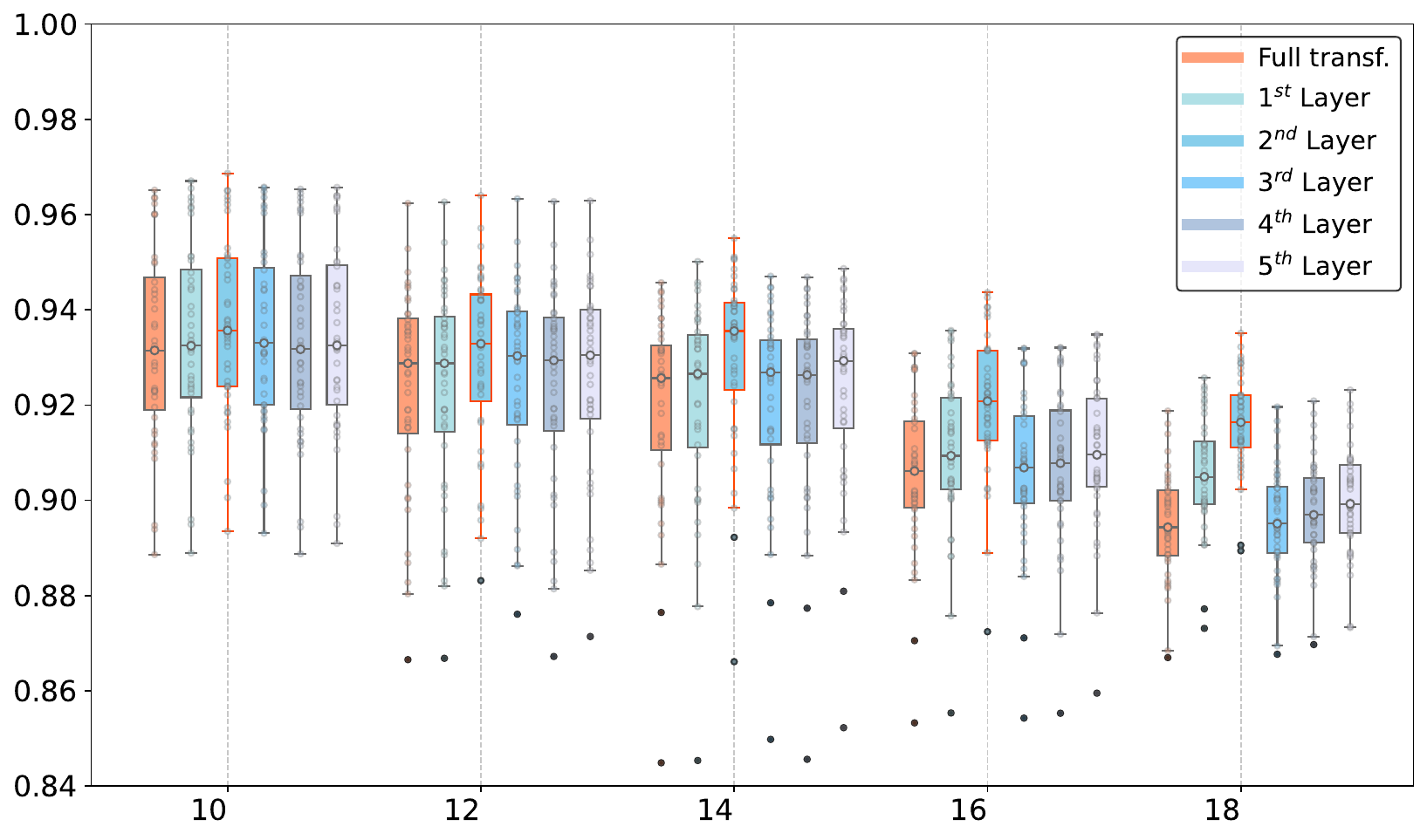}};
    \begin{scope}[x={(image.south east)}, y={(image.north west)}]
      \node at (0.53, -0.05) {\normalsize $N_2$}; 
      \node[rotate=90] at (-0.05, 0.52) {\Large $r$}; 
    \end{scope}
  \end{tikzpicture}
    \caption{Approximation ratio $r$ for full transfer, as well as for optimizing each layer individually for 40 instances of $N_{2}$-node acceptor graphs. The boxes represent the interquartile distance between the first and third quartile, covering $75\%$ of the data. The gray surrounded circle indicates the median, while all the remaining dots are the original approximation ratios for the $40$ graph samples. Points lying outside the box are outliers that are not covered within the interquartile distance.}
    \label{BOXPLOT_OVERn}
\end{figure*}

\begin{figure}[t!]
    \centering
    \begin{tikzpicture}

    \node[anchor=south west,inner sep=0] (image) at (0,0) 
        {\includegraphics[width=0.4\textwidth]{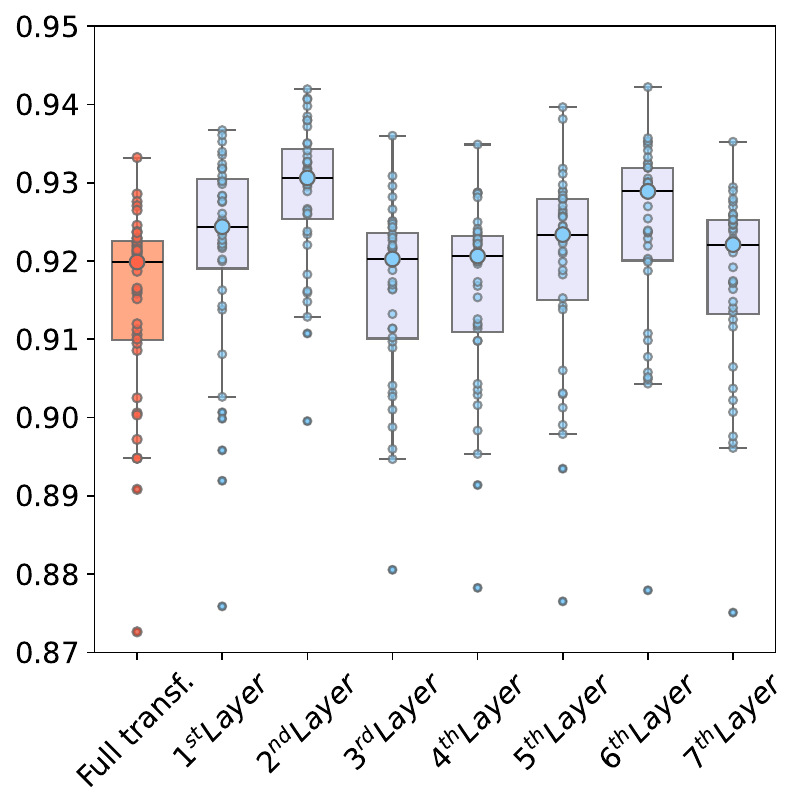}};
        \label{fig:single_layer_7lrs}
        \begin{scope}[x={(image.south east)}, y={(image.north west)}]
      \node at (0.51, -0.04) {\normalsize $\mathrm{Optimization\; scheme}$}; 
      \node[rotate=90] at (-0.05, 0.58) {\Large $r$}; 
    \end{scope}
    \end{tikzpicture}
    \caption{For $N_2 = 12$ and for QAOA circuits with $p=7$ layers, the approximation ratio $r$ for full parameter transfer as well as for individual optimization of each layer, calculated for $40$ acceptor graphs.}
    \label{fig:merged_single_layer}
\end{figure}

\begin{figure}[htbp]
    \centering
    \hspace{-0.3cm}
    \begin{tikzpicture}
        \node[anchor=south west, inner sep=0] (image) at (0,0)
            {\includegraphics[width=0.9\linewidth]{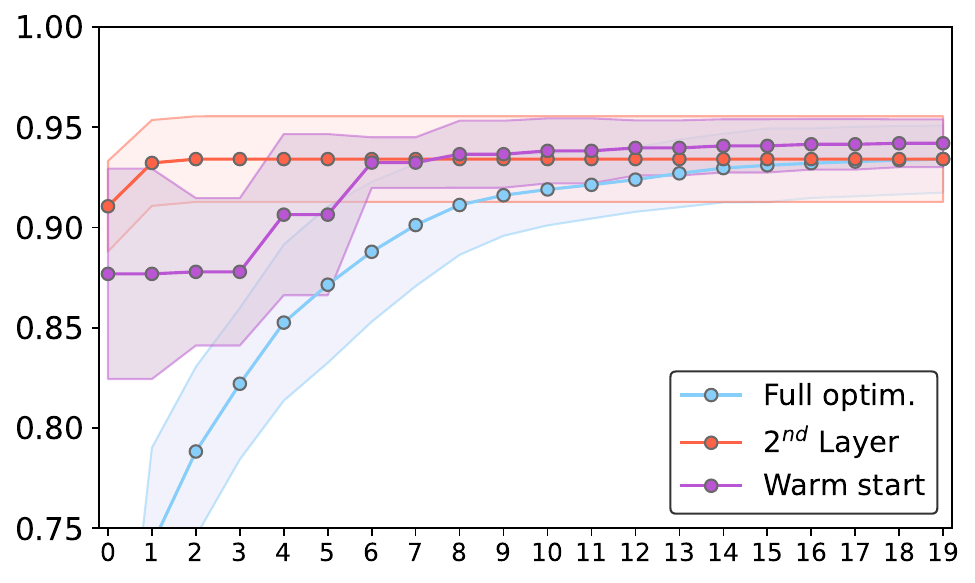}};
        \node at (4., -0.4) {\large $t$};
        \node[rotate=90] at (-0.38, 2.3) {\large $\frac{\langle H_c\rangle}{E_{\mathrm{min}}}$};
    \end{tikzpicture}
    \caption{Convergence of the ratio $\langle H_c\rangle/E_{\mathrm{min}}$ with increasing iteration steps $t$ for (\romannumeral 1) all layer optimization starting from random parameters (`Full optim.' blue curve), (\romannumeral 2) all layer optimization starting from transferred parameters (`Warm start' purple curve), and (\romannumeral 3) second layer optimization starting from transferred parameters (orange curve). The curves present the mean value while the shaded region indicates the standard deviation for 40 acceptor graph instances with $N_2 = 12$.}
    \label{optStepsTask}
\end{figure}

While the above studies imply that the parameter transfer results in an approximation ratio close to that obtained by self-optimization, this statement may not hold strongly when considering small-sized graphs or transfer between graphs with a large size difference. Indeed, as we observe from Ref.~\cite{akshay_2021}, $\{\gamma_{i}, \beta_{i}\}$ change significantly between small graphs with $N<10$ and larger graphs when $N$ is of the order of a few tens. The results of Ref.~\cite{sakai2024linearly} show that when transferring parameters between different graphs with the same $N$, a high difference in edge probability between them may result in sampling the wrong solution state for the acceptor graph. The study in Ref.~\cite{montanezbarrera2024} shows a poor transferability between different COPs for higher values of $N$. These observations suggest that, starting from the transferred parameters, an additional optimization should be performed to fine-tune them to achieve the optimal solution. Due to warm start, the time required for additional optimization is less than the time for self-optimization. However, it can be further reduced if we selectively fine-tune the parameters corresponding to only some of the layers instead of re-optimizing all layers. We schematically present the `layer-selective transfer learning' scheme in Fig. \ref{qaoa_schematic}. In the following section, we describe our methodologies and numerical observations when studying this layer-selective optimization of QAOA for the Max-Cut problem.

\section{Results}
\label{results}

We compare the following two optimization schemes.
\begin{enumerate}
    \item Full transfer: We transfer the optimized Max-Cut QAOA parameters for a fixed instance of a donor graph $G_1$ with $N_{1}$ nodes to the QAOA circuits for several acceptor graphs $G_2$ with $N_{2}$ nodes, and calculate the approximation ratios obtained for the latter.
    \item $n$-layer optimization with transferred initial parameters: In the scheme we propose, following the full transfer of parameters, we use them as the initial parameters and we further optimize only $n (<p)$ number of layers.
\end{enumerate}
We conduct each experiment using a five-layer QAOA circuit. As $G_{1}$, we use a connected and unweighted eight-node donor graph generated from the Erdos-Renyi distribution using a fixed seed, with edge probability $\mathcal{P}_{d}=0.6$, and perform self-optimization starting from random initial parameters. An eight-node graph represents a moderate size within the range of nodes chosen for our numerical simulations. As such, it may help optimize smaller graphs by exploring a broader range of complex paths in the parameter space.
For higher-node graphs, however, this initial configuration may lack sufficient complexity, which is why we incorporate an additional optimization step. For acceptor graphs, we randomly generate 40 connected and unweighted graphs from the same distribution and with same edge probability $\mathcal{P}_{a}=0.6$ as the donor graph. To measure the optimization time $\tau$, we use the number of iterations required to obtain the desired convergence of the cost function. In all cases, we stop the optimization cycle when the change in the cost function is less than $10^{-4}$. In some cases, we present a comparison of the approximation ratio obtained from the above two schemes with the warm-start optimization scheme. The latter refers to the scenario in which all layers are optimized starting from the transferred parameters.

We present a pseudo-algorithm of the selective optimization scheme in in Alg.\ref{our_algo}. From Erdos-Renyi distribution $\mathcal{G}$, we randomly generate a donor graph $G_1$ with $N_1$ nodes and set of edges $E_1$, and an acceptor graph $G_2$ with $N_2$ nodes and set of edges $E_2$, with $N_1 \leq N_2$. We set the number of maximum iterations $T$ as a large integer, and a tolerance tol$= 10^{-4}$. A $p$-layer QAOA circuit is optimized to find the optimal parameters for $G_1$. These parameters constitute the initial parameters for the acceptor graph $G_2$. Subsequently, a subset of $n$ layers is selected and optimized for $G_2$, keeping the rest of the parameters untouched during this process. When the difference between the values of the cost function for two consecutive steps is less than $tol$, and it repeats for $s=3$ iterations, we stop the optimization cycle.
We use Pennylane~\cite{pennylane} quantum simulator to build the circuits. We use the Adagrad optimizer from Optax~\cite{deepmind2020jax}, a gradient-based iterative optimizer. The entire simulation of the process has been conducted using JAX~\cite{deepmind2020jax} for just-in-time compilation and fast vector calculation.

To begin with, we examine how the approximation ratio improves when one of the five layers is optimized after parameter transfer. Figure \ref{BOXPLOT_OVERn} shows the distribution of approximation ratios for individually optimizing all five layers for 40 randomly generated acceptor graphs with even number of nodes varying from 10 to 18. We did not use a higher number of samples since that requires a significantly high simulation time for large graphs. 
However, $40$ samples are not sufficient to realize normally distributed data. Therefore, to represent the aggregate behavior of the approximation ratios, we use the median and the interquartile distance, the latter being the difference between the first and the third quartile of our samples that represent $75\%$ of the entire data samples. We compare them with the distribution of approximation ratios obtained using the full transfer of parameters without any further optimization. In all cases, we observe that training the second layer results in the highest improvement in the approximation ratio. We note that, when observing each acceptor graph individually, the above observation holds for the majority of them, albeit not for all. As expected, the median approximation ratio decreases with increasing $N_{2}$, demonstrating a diminishing transferability of parameters with an increasing size difference between the donor and acceptor graphs. Therefore, the improvement due to selective optimization becomes more pronounced for graphs with larger $N_{2}$.

In Fig. \ref{fig:merged_single_layer}, we show the approximation ratio obtained from single-layer optimization when using $p=7$ layers for the same set of donor and acceptor graphs as in Fig. \ref{BOXPLOT_OVERn}. In this case as well, optimizing the second layer alone yields the greatest improvement. We also verify that this observation holds true when changing the donor graph instance, keeping the node number and edge probability unchanged.

A major benefit of selective optimization is the low optimization time. In Fig. \ref{optStepsTask}, we show how fast the ratio $\langle H_{c}\rangle/E_{\mathrm{min}}$ improves with increasing iteration for all-layer-optimization starting from random initial parameters, warm-start optimization starting from transferred initial parameters, and optimizing the second layer alone starting from transferred initial parameters. We show the overall trend of $\langle H_{c}\rangle/ E_{\mathrm{min}}$ over $40$ random acceptor graphs. While the selective optimization scheme saturates after the second step on average, the warm-start scheme takes about ten steps to saturate. The first scheme does not reach a saturation point for $20$ iteration steps, indicating the long time required for optimization with random initial parameters.

\begin{figure}[t]
\hspace{-0.2cm}
    \centering
    \begin{tikzpicture}
        \node[anchor=south west, inner sep=0] (image) at (0,0)
            {\includegraphics[width=0.95\linewidth]{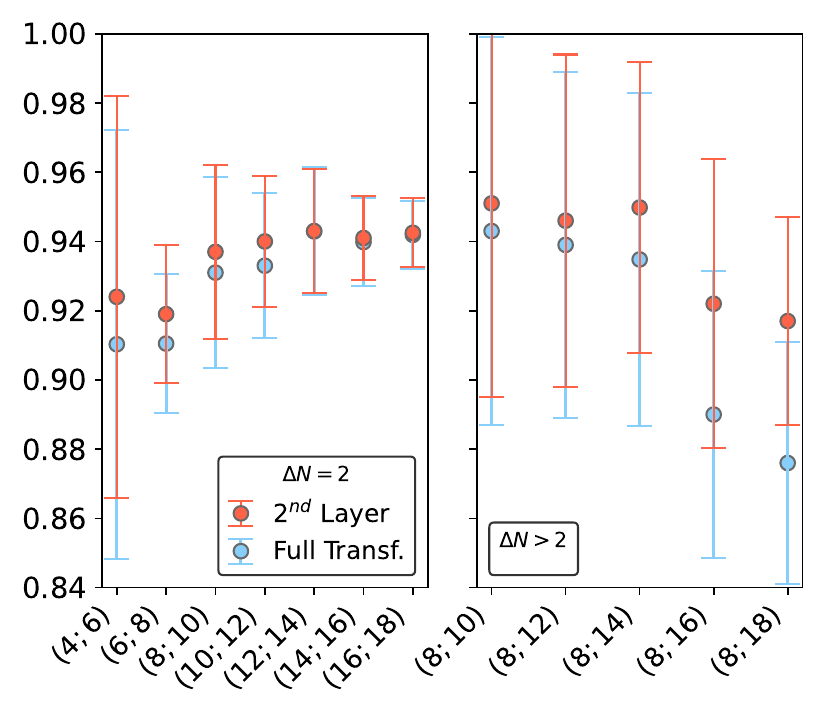}};
        \node at (2.2, 0.01){(a)};
        \node at (5.87, 0.01){(b)};
        \node at (4, -0.4) {$(N_{1};N_{2})$};
        \node[rotate=90] at (-0.2, 3.73) {\large$r$};
    \end{tikzpicture}
    \caption{(a) The approximation ratio $r$ for transfer of optimal parameters without (blue) and with (red) a further optimization of the second layer for $\Delta N = N_{2}-N_{1}= 2$. (b) The approximation ratio $r$ for $\Delta N > 2$ with a fixed $N_{1}=8$. For both subfigures, the circular points represent the median of the approximation ratio for $40$ different acceptor graph instances, while the vertical lines represent the corresponding interquartile distance between first and third quartile.}
    \label{scalab}
\end{figure}

Previous studies have investigated the scaling behavior of parameter transferability with respect to system size for Max-Cut problem as well as for the Sherrington Kirkpatrick model~\cite{Farhi2022, akshay_2021}. These studies found that the optimal QAOA parameters exhibited a converging behavior with respect to increasing size of the Max-Cut instances. Therefore, it is expected that, when both donor and acceptor graphs have a large number of nodes, parameter transfer will be more successful, or in other words, an additional optimization will result in a minor improvement. We examine this behavior in the context of selective optimization. We vary the number of nodes $N_1$ and $N_2$ of the donor and acceptor graphs, keeping the size difference $\Delta N = \vert N_{2}-N_{1}\vert$ the same in all cases. Thus, for $\Delta N=2$, we have a set of donor-acceptor pairs $\{(N_{1}; N_{1}+2)\}$ where $6 \leq N_{1} \leq 16$. Figure \ref{scalab}(a) shows the approximation ratios for full transfer and the second layer optimization for these donor-acceptor pairs. With the increase of $N_{1}$ and $N_{2}$, the approximation ratio corresponding to full transfer increases and then tends to saturate. On the other hand, the improvement due to an additional optimization diminishes and gradually becomes negligible. These observations indicate better transferability between large graphs. Figure \ref{scalab}(b) shows the approximation ratio for full transfer and second layer optimization for a fixed donor graph and increasing size of acceptor graphs. With increasing size difference $\Delta N$, the approximation ratio for full transfer decreases, showing poor transferability. As a result, the improvement due to selective optimization increases.

\begin{figure}[t!]
    \hspace{-0.48cm}
    \begin{tikzpicture}
        \node[anchor=south west, inner sep=0] (image) at (0,0)
            {\includegraphics[scale=0.45]{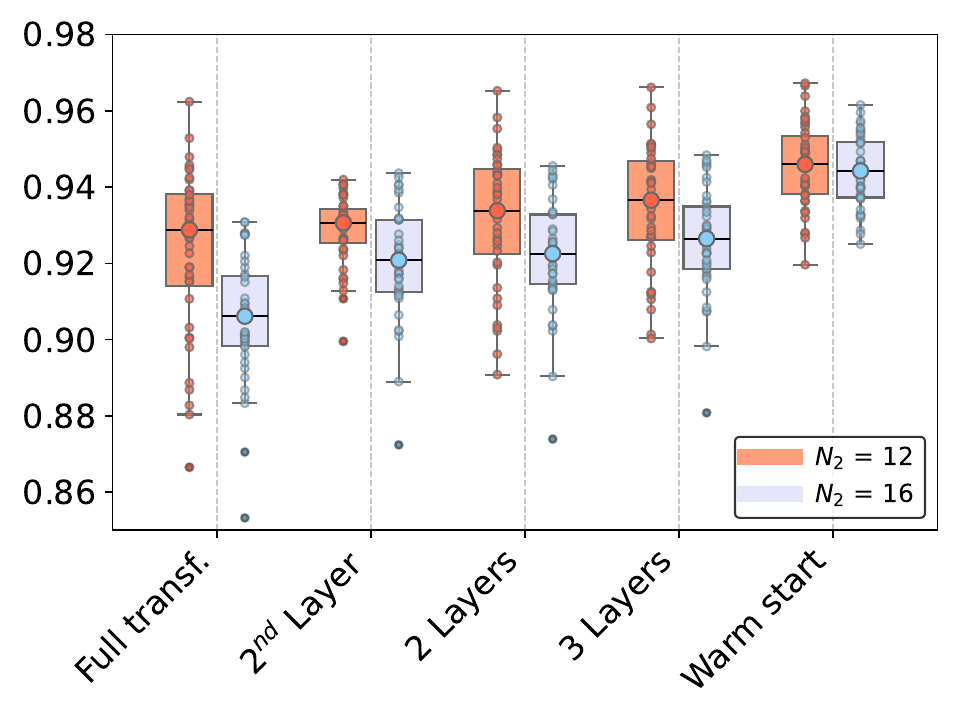}};
        \node at (4., -0.2) {\normalsize $\mathrm{Optimization\; scheme}$};
        \node[rotate=90] at (-0.3, 3.48) {\large $r$};
    \end{tikzpicture}

    \caption{Approximation ratios $r$ obtained from full parameter transfer, as well as from optimizing the second layer alone, the first two layers, the first three layers, and optimizing all layers for $N_{2}=12, 16$ acceptor graph instances.}
    \label{multiple_scheme_compare}
\end{figure}

\begin{figure*}[t]
\hspace{0.5cm}
\centering
\begin{tikzpicture}
    \node[anchor=south west,inner sep=0] (image) at (0,0)
        {\includegraphics[width=1.05\textwidth]{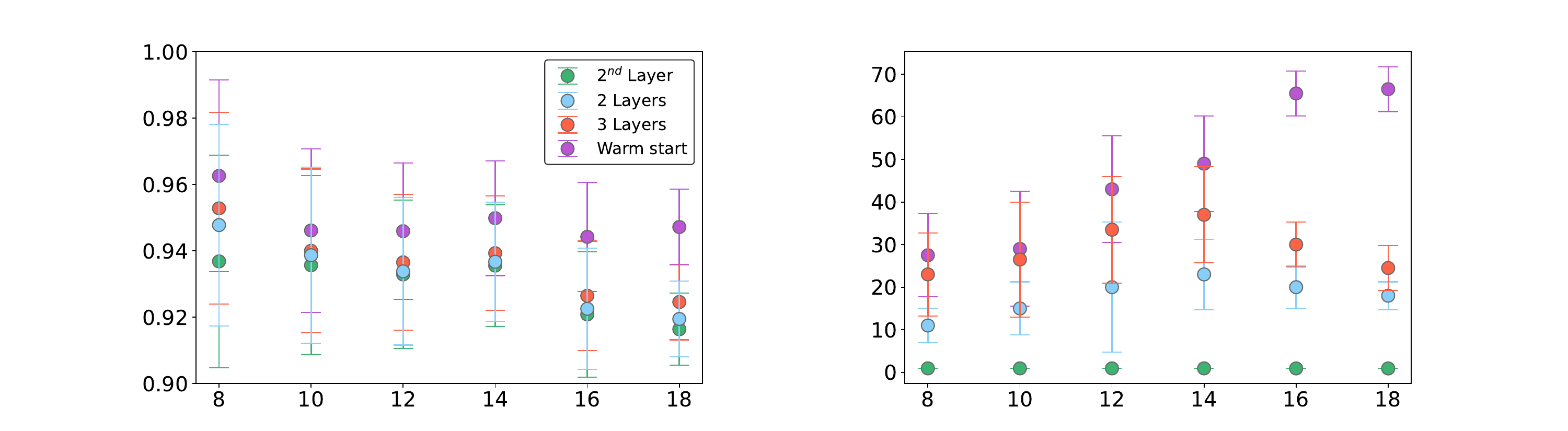}};
    \begin{scope}[x={(image.south east)}, y={(image.north west)}]
      \node at (0.29, -0.15){(a)};
      \node at (0.29, -0.03) {\normalsize $N_2$};  
      \node[rotate=90] at (0.07, 0.5) {\Large $r$}; 
      \node at (0.74, -0.03) 
      {\normalsize $N_2$};
      \node at (0.74, -0.15){(b)};
      \node[rotate=90] at (0.53, 0.5) {\Large $\tau$}; 
    \end{scope}
\end{tikzpicture}
\caption{(a) The approximation ratio $r$ and (b) the optimization time $\tau$ for different optimization schemes for varying number of nodes. The circular points represent the median while the vertical lines represent the interquartile distance between the first and third quartile of approximation ratios for 40 acceptor graphs.}
\label{custom_metric_plot}
\end{figure*}

\begin{figure*}[t]
\hspace{-1cm}
\hspace{0.5cm}
\centering
\begin{tikzpicture}
    \node[anchor=south west,inner sep=0] (image) at (0,0)
        {\includegraphics[width=0.8\textwidth]{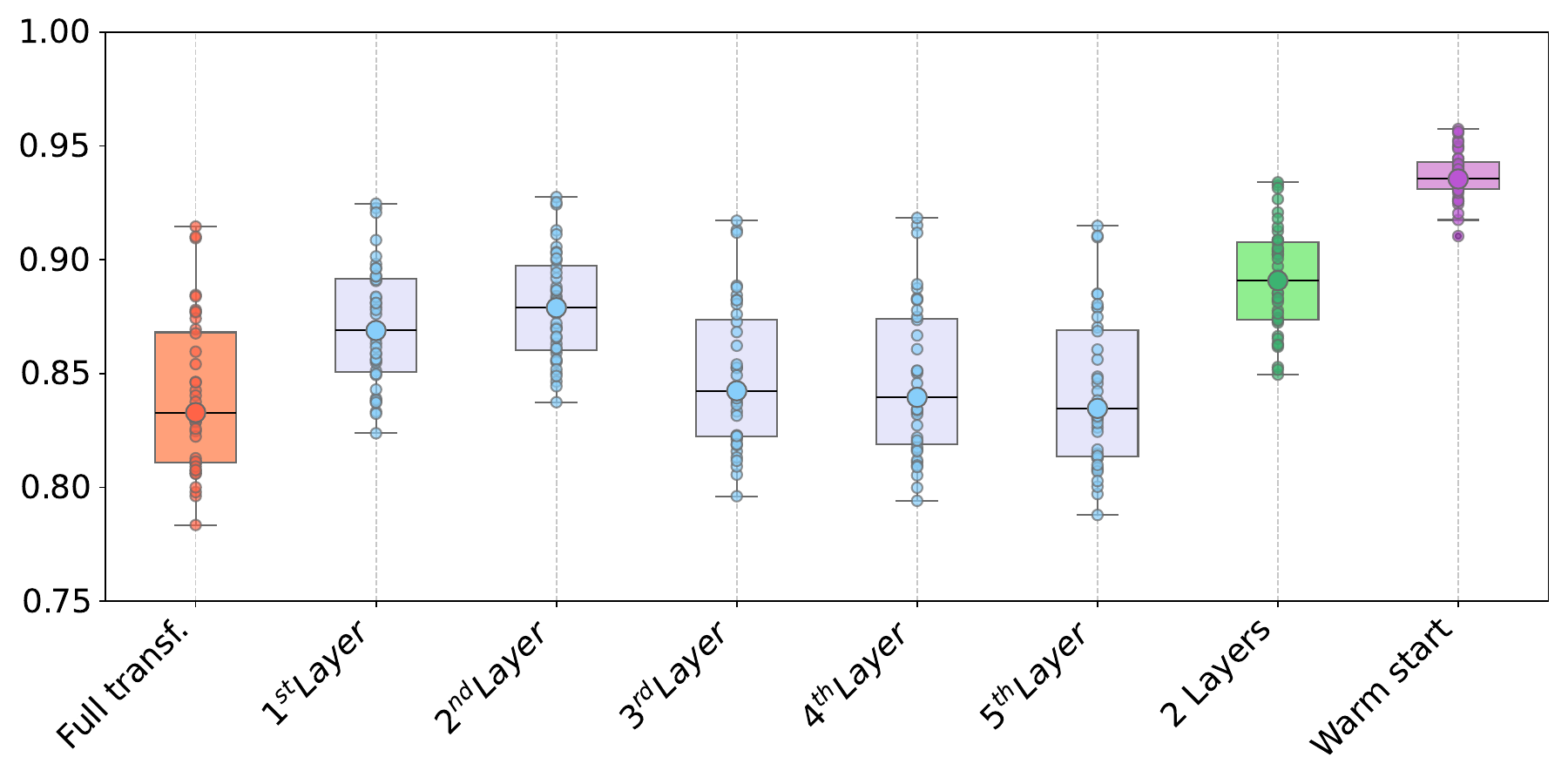}};
    
    \begin{scope}[x={(image.south east)}, y={(image.north west)}]
      \node at (0.5, -0.05) {   \normalsize $\mathrm{Optimization\; scheme}$};  
      \node[rotate=90] at (-0.02, 0.6) {\Large $r$}; 
    \end{scope}
\end{tikzpicture}
\caption{Median and interquartile distance of the approximation ratios $r$ for different optimization schemes of weighted graphs with $N_2=12$. The donor graph corresponds to $N_1=8$.}
\label{weighted_comparison}
\end{figure*}

\begin{table*}
  \centering
\begin{tabular}{|p{1.cm}||p{2.5cm}|p{2.5cm}|p{2.5cm}|p{2.5cm}|p{2.5cm}| }
\hline
\multicolumn{6}{|c|}{$\langle H_c\rangle_{\mathrm{min}}$ for single-layer optimization} \\
\hline
$N$&$\mathcal{}{1}$& $\mathcal{}{2}$ &$\mathcal{}{3}$ & $\mathcal{}{4}$& $\mathcal{}{5}$\\
\hline
$10$&$-18.054$& $\mathbf{-18.078}$& $-18.074$& $-18.077$& $-18.051$\\
\hline
$12$&$-26.546$& $\mathbf{-26.685}$& $-26.580$& $-26.568$& $-26.646$\\
\hline
$14$&$-35.973$& $\mathbf{-36.576}$& $-36.056$& $-36.060$& $-36.116$\\
\hline
$16$&$-46.571$& $\mathbf{-47.362}$& $-46.540$& $-46.571$& $-46.662$\\
\hline
\end{tabular}
\caption{The minimum energy $\langle H_c \rangle_{\mathrm{min}}$ obtained from grid search in the $\{\gamma_{i}, \beta_{i}\}$ plane of layer $i$ ($i=1, 2, \ . \ . \ ., \ 5$) of a five-layer QAOA circuit, while keeping the other parameters fixed at the values transferred from the eight-node donor graph. The results correspond to optimizing one particular instance of an acceptor graph for each node number. The minimum energy for each node number is highlighted in boldface characters.}
\label{table_energy}
\end{table*}

Next, we investigate the selective optimization of two or three layers among the total of five layers. Motivated by our observations in Fig. \ref{BOXPLOT_OVERn}, we choose to optimize the first two or three layers exclusively. A comparative analysis of this optimization scheme with full transfer and warm-start optimization is presented in Fig. \ref{multiple_scheme_compare}. 
It is clear that optimizing a subset of layers fails to reach an approximation ratio close to that obtained from warm-start optimization of all layers. We also observed that for large graphs (e.g. $N_{2}=18$), optimizing the second layer alone results in a pronounced improvement in the approximation ratio compared to full transfer. However, the additional improvement resulting from optimizing two or three layers is not substantial for both $N_2=12, 18$.

We further investigate the trade-off between the approximation ratio $r$ and the optimization time $\tau$. In Fig. \ref{custom_metric_plot}(a) and (b), we show the median values of these two quantities with varying numbers of nodes. These figures demonstrate that when selectively optimizing more than one layer, the gain in approximation ratio is very small, while the number of iterations needed to obtain the desired convergence is very high. On the other hand, optimizing the second layer alone requires only a few iterations. 
For $N_2=16, 18$, the cost function undergoes small oscillations near convergence, due to which the warm start optimization requires a significantly high $\tau$. However, for these large graphs, $\tau$ corresponding to two- or three-layers optimization is relatively low, and the improvement in $r$ over the single-layer optimization is higher than the cases when $N_{2}< 16$.
Therefore, the efficiency of 2-layer and 3-layer optimization begins to surpass the all-layer optimization only when the number of nodes is sufficiently high. We could not use $N_{2}>18$ since the simulation time increases significantly.

Finally, we investigate the selective optimization scheme for weighted graphs. The edge weights are randomly generated for donor and acceptor graphs. In Fig. \ref{weighted_comparison}, we compare individual single-layer optimization as well as two- and three-layer optimization with full parameter transfer and warm-start optimization. We again observe the highest approximation ratio for optimizing the second layer. Also, optimizing the first two layers results in little improvement over optimizing the second layer alone. However, all transfer learning schemes have significantly lower approximation ratios compared to all-layer optimization with warm-start. Thus, for weighted graphs, sampling the correct solution may necessarily require optimizing all layers following the parameter transfer.

To gain a clearer understanding of why the second layer gives the best approximation ratio for selective optimization, we initialize the QAOA circuit with transferred parameters, and then perform a grid search for $\langle H_{c} \rangle_{\mathrm{min}}$ in the $\{\gamma_{i}, \beta_{i}\}$ plane in the range $[-1, 1]$ for $i=1, 2, .., 5$, while keeping the other parameters unchanged. The data presented in Table \ref{table_energy} shows that $\langle H_{c} \rangle_{\mathrm{min}}$ is lower in $\{\gamma_{2}, \beta_{2}\}$ plane compared to the four other energy planes, which explains the best value of $r$ obtained for the second layer. Therefore, it is clear that our observed trend arises from the nature of the loss landscape. For any optimization problem, optimizing a subset of all free parameters implies searching the minima of the loss landscape along a constrained trajectory. Therefore, when optimizing different sets of parameters of QAOA, one may encounter a better minima along a particular trajectory, compared to others.
This can be better understood by obtaining an analytical expression of $\langle H_{c} \rangle$ in terms of $\{\gamma_{i}, \beta_{i}\}$ for Max-Cut problem. 
In Ref.~\cite{wang_2018} the authors analytically study the loss landscape for $p=1$ layer QAOA for Max-Cut graphs, which shows the dependence of the loss function on the degree of connectivity of the vertices. However, for QAOA with $p>1$ levels, an analytical study of the loss landscape is challenging and rigorous, and remains unexplored to date. Since our observations are based on five-layer and seven-layer QAOA circuits, a more clear and analytical reasoning behind our observed trend of layer-wise approximation ratio remains beyond the scope of this work.

\begin{figure}[t!]
\hspace{-0.3cm}
    \centering
    \begin{tikzpicture}
        \node[anchor=south west, inner sep=0] (image) at (0,0)
            {\includegraphics[width=0.96\linewidth]{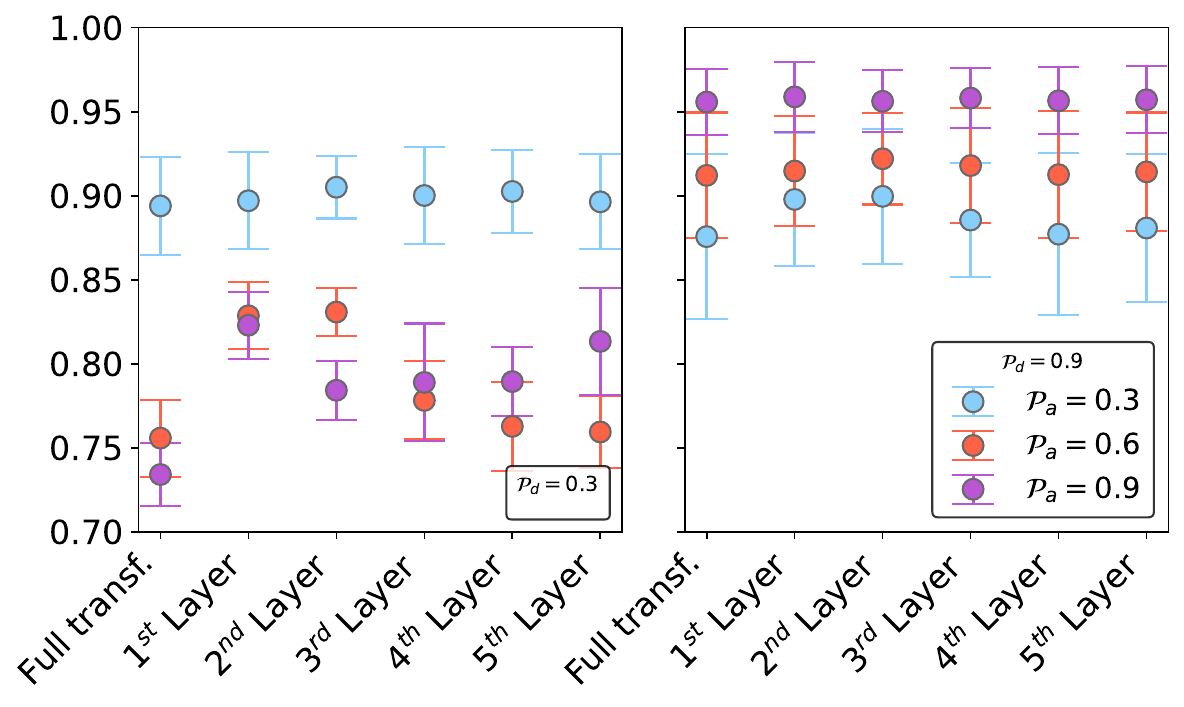}};
        \large \node[rotate=90] at (-0.2, 2.7) {$r$};
        \node at (4.14, -0.8) {\normalsize$\mathrm{Optimization\; scheme}$};
        \node at (2.4, -0.2){\small (a)};
        \node at (5.9, -0.2){\small (b)};
    \end{tikzpicture}
    \caption{A comparison between approximation ratios $r$ obtained from donor graphs with edge probability $\mathcal{P}_d = 0.3$ (a) and $\mathcal{P}_d = 0.9$ (b). The different colors of the sample points represent acceptor graphs with edge probability $\mathcal{P}_a = [0.3, \ 0.6, \ 0.9]$ to which we transferred the optimal donor parameter. Each circular point represent the median across $40$ instances and the vertical lines the interquartile distance between $1^{st}$ and $3^{rd}$ quartile.}
    \label{prob_task}
\end{figure}

However, we numerically investigate whether our observed trend is dependent on the graph connectivity by analyzing the approximation ratio under selective optimization for varying edge probability of the donor and acceptor graphs. We show the results in Fig. \ref{prob_task} for $8$-node donor graphs and $12$-node acceptor graphs. For donor graphs with edge probability $0.3$, transferring parameters to acceptor graphs with $\mathcal{P}_{a}=0.3, 0.6$ yields the highest approximation ratio for selective optimization of the second layer. In contrast, for acceptor graphs with $\mathcal{P}_{a}=0.9$, the first layer achieves the best performance, while the second layer performs worst. For donor graphs with $\mathcal{P}_{d}=0.9$, transfer to acceptor graphs with lower connectivity ($\mathcal{P}_{a}=0.3, 0.6$) again shows optimal performance at the second layer. Quite surprisingly, in this case the first layer performs best when the acceptor graphs have the same probability $\mathcal{P}_{a}=0.9$ as the donor graphs. In Table \ref{table_energy_prob}, we show the layer-wise approximation ratio for varying $\mathcal{P}_{d}$, which shows that the first or the second layer performs best among all layers. Therefore, although in the majority of our observed cases the second layer achieves the best performance under selective optimization, the results indicate that depending on the $\mathcal{P}_{d}$, the optimum lies in either the first or the second layer.

\begin{table*}
  \centering
\begin{tabular}{|p{1.cm}|p{1.cm}||p{2.5cm}|p{2.5cm}|p{2.5cm}|p{2.5cm}|p{2.5cm}| }
\hline
\multicolumn{7}{|c|}{$\langle H_c\rangle_{\mathrm{min}}$ for single-layer optimization} \\
\hline
$\mathcal{P}_{d}$&$\mathcal{P}_{a}$&$\mathcal{}{1}$& $\mathcal{}{2}$ &$\mathcal{}{3}$ & $\mathcal{}{4}$& $\mathcal{}{5}$\\
\hline
$0.3$&$0.9$&$\mathbf{-30.662}$& $-30.600$& $-30.104$& $-29.910$& $-29.848$\\
\hline
$0.5$&$0.9$&$-31.992$& $\mathbf{-33.079}$& $-30.528$& $-30.018$& $-30.525$\\
\hline
$0.7$&$0.9$&$\mathbf{-34.132}$& $-34.074$& $-33.924$& $-33.928$& $-33.644$\\
\hline
$0.9$&$0.9$&$\mathbf{-34.538}$& $-34.393$& $-34.466$& $-34.430$& $-34.449$\\
\hline
\end{tabular}
\caption{The minimum energy $\langle H_c \rangle_{\mathrm{min}}$ obtained from grid search in the $\{\gamma_{i}, \beta_{i}\}$ plane of layer $i$ ($i=1, 2, \ . \ . \ ., \ 5$) of a five-layer QAOA circuit, when the initial parameters are transferred from eight-node donor graphs with varying edge probabilities $\mathcal{P}_{d}$ to a fixed instance of a twelve-node acceptor graph with edge probability $\mathcal{P}_{a}=0.9$. The minimum energy for each donor graph is highlighted in boldface characters.}
\label{table_energy_prob}
\end{table*}

\section{Conclusion}
\label{conclusion}
\vspace{-0.3cm}
The QAOA has established itself as a promising algorithm for solving COPs such as the Max-Cut problem. The costly optimization process of a high-depth QAOA circuit for large problem sizes can be bypassed by using the good transferability of parameters between different problem instances. We have proposed an approach to layer-selective optimization of QAOA parameters after the parameter transfer. The optimal QAOA parameters of a donor graph are used to initialize the QAOA circuit for larger acceptor graphs. Following this, we proposed to optimize a subset of layers, rather than all the layers. We have showed that this approach takes drastically lower optimization time compared to optimizing all layers. We found that the selective optimization becomes more useful when the donor and acceptor graphs have a large difference in size.
Our study also highlighted the different contributions arising from each single layer of the circuit, and showed that the optimization of a specific layer can give higher or lower approximation ratio values, implying different significance of individual layers. In particular, the numerical results suggest that optimizing the first or the second layer alone corresponds to the highest trade-off between approximation quality and the required optimization time. These observations can lead to new insights toward understanding how the QAOA works and how to set up efficient and powerful QAOA-based algorithms. 
The next step could be to test the validity of our results on real quantum processors and on larger graphs to open new ways to efficiently implement a pipeline to assess COPs via quantum machine learning.

\acknowledgements

This work was supported by the European Union’s Research and Innovation Programme Horizon Europe G.A. no. 101070546 (MUQUABIS), by the European Defence Agency under the project Q-LAMPS Contract No. B PRJ- RT-989, by the PNRR MUR project PE0000023-NQSTI, and by the MUR Progetti di Ricerca di Rilevante Interesse Nazionale (PRIN) Bando 2022, Project No. 20227HSE83 – ThAI-MIA funded by the European Union - Next Generation EU.


\appendix


\bibliography{main}

@inproceedings{BB84,
  address = {India},
  author = {Bennett, C. H. and Brassard, G.},
  booktitle = {Proceedings of IEEE International Conference on Computers, Systems, and Signal Processing},
  location = {Bangalore},
  pages = 175,
  title = {{Quantum cryptography: Public key distribution and coin tossing}},
  year = 1984
}

@article{shor,
author = {Shor, Peter W.},
title = {Polynomial-Time Algorithms for Prime Factorization and Discrete Logarithms on a Quantum Computer},
journal = {SIAM Journal on Computing},
volume = {26},
number = {5},
pages = {1484-1509},
year = {1997},
doi = {10.1137/S0097539795293172},
URL = { https://doi.org/10.1137/S0097539795293172},
eprint = {https://doi.org/10.1137/S0097539795293172}
}

@inproceedings{grover1996,
author = {Grover, Lov K.},
title = {A fast quantum mechanical algorithm for database search},
year = {1996},
isbn = {0897917855},
publisher = {Association for Computing Machinery},
address = {New York, NY, USA},
url = {https://doi.org/10.1145/237814.237866},
doi = {10.1145/237814.237866},
booktitle = {Proceedings of the Twenty-Eighth Annual ACM Symposium on Theory of Computing},
pages = {212–219},
numpages = {8},
location = {Philadelphia, Pennsylvania, USA},
series = {STOC '96}
}

@article{teleport_1993,
  title = {Teleporting an unknown quantum state via dual classical and {E}instein-{P}odolsky-{R}osen channels},
  author = {Bennett, Charles H. and Brassard, Gilles and Cr\'epeau, Claude and Jozsa, Richard and Peres, Asher and Wootters, William K.},
  journal = {Phys. Rev. Lett.},
  volume = {70},
  issue = {13},
  pages = {1895--1899},
  numpages = {0},
  year = {1993},
  month = {Mar},
  publisher = {American Physical Society},
  doi = {10.1103/PhysRevLett.70.1895},
  url = {https://link.aps.org/doi/10.1103/PhysRevLett.70.1895}
}

@misc{farhi2014qaoa,
      title={A Quantum Approximate Optimization Algorithm}, 
      author={Edward Farhi and Jeffrey Goldstone and Sam Gutmann},
      year={2014},
      eprint={1411.4028},
      archivePrefix={arXiv},
      primaryClass={quant-ph},
      url={https://arxiv.org/abs/1411.4028}, 
}

@article{Hadfield_2019,
   title={From the Quantum Approximate Optimization Algorithm to a Quantum Alternating Operator Ansatz},
   volume={12},
   ISSN={1999-4893},
   url={http://dx.doi.org/10.3390/a12020034},
   DOI={10.3390/a12020034},
   number={2},
   journal={Algorithms},
   publisher={MDPI AG},
   author={Hadfield, Stuart and Wang, Zhihui and O’Gorman, Bryan and Rieffel, Eleanor G. and Venturelli, Davide and Biswas, Rupak},
   year={2019},
   month=feb, pages={34} }

@misc{brandao2018,
      title={For Fixed Control Parameters the Quantum Approximate Optimization Algorithm's Objective Function Value Concentrates for Typical Instances}, 
      author={Fernando G. S. L. Brandao and Michael Broughton and Edward Farhi and Sam Gutmann and Hartmut Neven},
      year={2018},
      eprint={1812.04170},
      archivePrefix={arXiv},
      primaryClass={quant-ph},
      url={https://arxiv.org/abs/1812.04170}, 
}

@ARTICLE{galda2023similarity,
  
AUTHOR={Galda, Alexey  and Gupta, Eesh  and Falla, Jose  and Liu, Xiaoyuan  and Lykov, Danylo  and Alexeev, Yuri  and Safro, Ilya },
         
TITLE={Similarity-based parameter transferability in the quantum approximate optimization algorithm},
        
JOURNAL={Frontiers in Quantum Science and Technology},
        
VOLUME={Volume 2 - 2023},

YEAR={2023},

URL={https://www.frontiersin.org/journals/quantum-science-and-technology/articles/10.3389/frqst.2023.1200975},

DOI={10.3389/frqst.2023.1200975},

ISSN={2813-2181}}

@article{shaydulin_2023,
author = {Shaydulin, Ruslan and Lotshaw, Phillip C. and Larson, Jeffrey and Ostrowski, James and Humble, Travis S.},
title = {Parameter Transfer for Quantum Approximate Optimization of Weighted {M}ax{C}ut},
year = {2023},
issue_date = {September 2023},
publisher = {Association for Computing Machinery},
address = {New York, NY, USA},
volume = {4},
number = {3},
url = {https://doi.org/10.1145/3584706},
doi = {10.1145/3584706},
journal = {ACM Transactions on Quantum Computing},
month = apr,
articleno = {19},
numpages = {15}
}

@article{Farhi2022,
  doi = {10.22331/q-2022-07-07-759},
  url = {https://doi.org/10.22331/q-2022-07-07-759},
  title = {The {Q}uantum {A}pproximate {O}ptimization {A}lgorithm and the {S}herrington-{K}irkpatrick {M}odel at {I}nfinite {S}ize},
  author = {Farhi, Edward and Goldstone, Jeffrey and Gutmann, Sam and Zhou, Leo},
  journal = {{Quantum}},
  issn = {2521-327X},
  publisher = {{Verein zur F{\"{o}}rderung des Open Access Publizierens in den Quantenwissenschaften}},
  volume = {6},
  pages = {759},
  month = jul,
  year = {2022}
}

@INPROCEEDINGS{galda_2021,
  author={Galda, Alexey and Liu, Xiaoyuan and Lykov, Danylo and Alexeev, Yuri and Safro, Ilya},
  booktitle={2021 IEEE International Conference on Quantum Computing and Engineering (QCE)}, 
  title={Transferability of optimal QAOA parameters between random graphs}, 
  year={2021},
  volume={},
  number={},
  pages={171-180},
  doi={10.1109/QCE52317.2021.00034}}

@misc{montanezbarrera2024,
      title={Transfer learning of optimal QAOA parameters in combinatorial optimization}, 
      author={J. A. Montanez-Barrera and Dennis Willsch and Kristel Michielsen},
      year={2024},
      eprint={2402.05549},
      archivePrefix={arXiv},
      primaryClass={quant-ph},
      url={https://arxiv.org/abs/2402.05549}, 
}

@misc{farhi2019quantumsupremacyquantumapproximate,
      title={Quantum Supremacy through the Quantum Approximate Optimization Algorithm}, 
      author={Edward Farhi and Aram W Harrow},
      year={2019},
      eprint={1602.07674},
      archivePrefix={arXiv},
      primaryClass={quant-ph},
      url={https://arxiv.org/abs/1602.07674}, 
}

@article{peruzzoVQA2014,
	author = {Peruzzo, Alberto and McClean, Jarrod and Shadbolt, Peter and Yung, Man-Hong and Zhou, Xiao-Qi and Love, Peter J. and Aspuru-Guzik, Al{\'a}n and O'Brien, Jeremy L.},
	date = {2014/07/23},
	doi = {10.1038/ncomms5213},
	id = {Peruzzo2014},
	isbn = {2041-1723},
	journal = {Nature Communications},
	number = {1},
	pages = {4213},
	title = {A variational eigenvalue solver on a photonic quantum processor},
	url = {https://doi.org/10.1038/ncomms5213},
	volume = {5},
	year = {2014}}

@article{McClean_2016,
doi = {10.1088/1367-2630/18/2/023023},
url = {https://dx.doi.org/10.1088/1367-2630/18/2/023023},
year = {2016},
month = {feb},
publisher = {IOP Publishing},
volume = {18},
number = {2},
pages = {023023},
author = {Jarrod R McClean and Jonathan Romero and Ryan Babbush and Alán Aspuru-Guzik},
title = {The theory of variational hybrid quantum-classical algorithms},
journal = {New Journal of Physics}
}

@ARTICLE{lucas_2014,

AUTHOR={Lucas, Andrew },

TITLE={Ising formulations of many {NP} problems},

JOURNAL={Frontiers in Physics},

VOLUME={2},

YEAR={2014},

URL={https://www.frontiersin.org/journals/physics/articles/10.3389/fphy.2014.00005},

DOI={10.3389/fphy.2014.00005},

ISSN={2296-424X}}

@article{mcclean_BP_2018,
	author = {McClean, Jarrod R. and Boixo, Sergio and Smelyanskiy, Vadim N. and Babbush, Ryan and Neven, Hartmut},
	date = {2018/11/16},
	doi = {10.1038/s41467-018-07090-4},
	id = {McClean2018},
	isbn = {2041-1723},
	journal = {Nature Communications},
	number = {1},
	pages = {4812},
	title = {Barren plateaus in quantum neural network training landscapes},
	url = {https://doi.org/10.1038/s41467-018-07090-4},
	volume = {9},
	year = {2018}}

@article{Sack2021quantumannealing,
  doi = {10.22331/q-2021-07-01-491},
  url = {https://doi.org/10.22331/q-2021-07-01-491},
  title = {Quantum annealing initialization of the quantum approximate optimization algorithm},
  author = {Sack, Stefan H. and Serbyn, Maksym},
  journal = {{Quantum}},
  issn = {2521-327X},
  publisher = {{Verein zur F{\"{o}}rderung des Open Access Publizierens in den Quantenwissenschaften}},
  volume = {5},
  pages = {491},
  month = jul,
  year = {2021}
}

@article{zhou_2020,
  title = {Quantum Approximate Optimization Algorithm: Performance, Mechanism, and Implementation on Near-Term Devices},
  author = {Zhou, Leo and Wang, Sheng-Tao and Choi, Soonwon and Pichler, Hannes and Lukin, Mikhail D.},
  journal = {Phys. Rev. X},
  volume = {10},
  issue = {2},
  pages = {021067},
  numpages = {23},
  year = {2020},
  month = {Jun},
  publisher = {American Physical Society},
  doi = {10.1103/PhysRevX.10.021067},
  url = {https://link.aps.org/doi/10.1103/PhysRevX.10.021067}
}

@article{Streif_2020,
doi = {10.1088/2058-9565/ab8c2b},
url = {https://dx.doi.org/10.1088/2058-9565/ab8c2b},
year = {2020},
month = {may},
publisher = {IOP Publishing},
volume = {5},
number = {3},
pages = {034008},
author = {Michael Streif and Martin Leib},
title = {Training the quantum approximate optimization algorithm without access to a quantum processing unit},
journal = {Quantum Science and Technology}
}

@INPROCEEDINGS{Lee_2021,
  author={Lee, Xinwei and Saito, Yoshiyuki and Cai, Dongsheng and Asai, Nobuyoshi},
  booktitle={2021 IEEE International Conference on Quantum Computing and Engineering (QCE)}, 
  title={Parameters Fixing Strategy for Quantum Approximate Optimization Algorithm}, 
  year={2021},
  volume={},
  number={},
  pages={10-16},
  doi={10.1109/QCE52317.2021.00016}}

@article{anschuetz_2022,
	author = {Anschuetz, Eric R. and Kiani, Bobak T.},
	date = {2022/12/15},
	doi = {10.1038/s41467-022-35364-5},
	id = {Anschuetz2022},
	isbn = {2041-1723},
	journal = {Nature Communications},
	number = {1},
	pages = {7760},
	title = {Quantum variational algorithms are swamped with traps},
	url = {https://doi.org/10.1038/s41467-022-35364-5},
	volume = {13},
	year = {2022}}

@misc{sakai2024linearly,
      title={Linearly simplified {QAOA} parameters and transferability}, 
      author={Ryo Sakai and Hiromichi Matsuyama and Wai-Hong Tam and Yu Yamashiro and Keisuke Fujii},
      year={2024},
      eprint={2405.00655},
      archivePrefix={arXiv},
      primaryClass={quant-ph},
      url={https://arxiv.org/abs/2405.00655}, 
}

@article{wang_2018,
  title = {Quantum approximate optimization algorithm for {M}ax{C}ut: A fermionic view},
  author = {Wang, Zhihui and Hadfield, Stuart and Jiang, Zhang and Rieffel, Eleanor G.},
  journal = {Phys. Rev. A},
  volume = {97},
  issue = {2},
  pages = {022304},
  numpages = {11},
  year = {2018},
  month = {Feb},
  publisher = {American Physical Society},
  doi = {10.1103/PhysRevA.97.022304},
  url = {https://link.aps.org/doi/10.1103/PhysRevA.97.022304}
}

@article{rajak_2023,
author = {Rajak, Atanu  and Suzuki, Sei  and Dutta, Amit  and Chakrabarti, Bikas K. },
title = {Quantum annealing: an overview},
journal = {Philosophical Transactions of the Royal Society A: Mathematical, Physical and Engineering Sciences},
volume = {381},
number = {2241},
pages = {20210417},
year = {2023},
doi = {10.1098/rsta.2021.0417}
}

@INPROCEEDINGS{eidenbenz_grover_2020,
  author={Bärtschi, Andreas and Eidenbenz, Stephan},
  booktitle={2020 IEEE International Conference on Quantum Computing and Engineering (QCE)}, 
  title={Grover Mixers for {QAOA}: Shifting Complexity from Mixer Design to State Preparation}, 
  year={2020},
  volume={},
  number={},
  pages={72-82},
  doi={10.1109/QCE49297.2020.00020}}

@article{wang_2020,
  title = {$XY$ mixers: Analytical and numerical results for the quantum alternating operator ansatz},
  author = {Wang, Zhihui and Rubin, Nicholas C. and Dominy, Jason M. and Rieffel, Eleanor G.},
  journal = {Phys. Rev. A},
  volume = {101},
  issue = {1},
  pages = {012320},
  numpages = {16},
  year = {2020},
  month = {Jan},
  publisher = {American Physical Society},
  doi = {10.1103/PhysRevA.101.012320},
  url = {https://link.aps.org/doi/10.1103/PhysRevA.101.012320}
}

@INPROCEEDINGS{cook_2020,
  author={Cook, Jeremy and Eidenbenz, Stephan and Bärtschi, Andreas},
  booktitle={2020 IEEE International Conference on Quantum Computing and Engineering (QCE)}, 
  title={The Quantum Alternating Operator Ansatz on Maximum k-Vertex Cover}, 
  year={2020},
  volume={},
  number={},
  pages={83-92},
  doi={10.1109/QCE49297.2020.00021}}

@Article{Lykov2023,
author={Lykov, Danylo
and Wurtz, Jonathan
and Poole, Cody
and Saffman, Mark
and Noel, Tom
and Alexeev, Yuri},
title={Sampling frequency thresholds for the quantum advantage of the quantum approximate optimization algorithm},
journal={npj Quantum Information},
year={2023},
month={Jul},
day={25},
volume={9},
number={1},
pages={73},
issn={2056-6387},
doi={10.1038/s41534-023-00718-4},
url={https://doi.org/10.1038/s41534-023-00718-4}
}

@article{shaydulin_2024,
author = {Ruslan Shaydulin  and Changhao Li  and Shouvanik Chakrabarti  and Matthew DeCross  and Dylan Herman  and Niraj Kumar  and Jeffrey Larson  and Danylo Lykov  and Pierre Minssen  and Yue Sun  and Yuri Alexeev  and Joan M. Dreiling  and John P. Gaebler  and Thomas M. Gatterman  and Justin A. Gerber  and Kevin Gilmore  and Dan Gresh  and Nathan Hewitt  and Chandler V. Horst  and Shaohan Hu  and Jacob Johansen  and Mitchell Matheny  and Tanner Mengle  and Michael Mills  and Steven A. Moses  and Brian Neyenhuis  and Peter Siegfried  and Romina Yalovetzky  and Marco Pistoia },
title = {Evidence of scaling advantage for the quantum approximate optimization algorithm on a classically intractable problem},
journal = {Science Advances},
volume = {10},
number = {22},
pages = {eadm6761},
year = {2024},
doi = {10.1126/sciadv.adm6761},
URL = {https://www.science.org/doi/abs/10.1126/sciadv.adm6761},
eprint = {https://www.science.org/doi/pdf/10.1126/sciadv.adm6761}}

@article{BLEKOS20241,
title = {A review on Quantum Approximate Optimization Algorithm and its variants},
journal = {Physics Reports},
volume = {1068},
pages = {1-66},
year = {2024},
issn = {0370-1573},
doi = {https://doi.org/10.1016/j.physrep.2024.03.002},
url = {https://www.sciencedirect.com/science/article/pii/S0370157324001078},
author = {Kostas Blekos and Dean Brand and Andrea Ceschini and Chiao-Hui Chou and Rui-Hao Li and Komal Pandya and Alessandro Summer}
}

@article{akshay_2021,
  title = {Parameter concentrations in quantum approximate optimization},
  author = {Akshay, V. and Rabinovich, D. and Campos, E. and Biamonte, J.},
  journal = {Phys. Rev. A},
  volume = {104},
  issue = {1},
  pages = {L010401},
  numpages = {6},
  year = {2021},
  month = {Jul},
  publisher = {American Physical Society},
  doi = {10.1103/PhysRevA.104.L010401},
  url = {https://link.aps.org/doi/10.1103/PhysRevA.104.L010401}
}

@Article{Lotshaw2021,
author={Lotshaw, Phillip C.
and Humble, Travis S.
and Herrman, Rebekah
and Ostrowski, James
and Siopsis, George},
title={Empirical performance bounds for quantum approximate optimization},
journal={Quantum Information Processing},
year={2021},
month={Nov},
day={25},
volume={20},
number={12},
pages={403},
issn={1573-1332},
doi={10.1007/s11128-021-03342-3},
url={https://doi.org/10.1007/s11128-021-03342-3}
}

@article{zhang2023review,
  title={A review on learning to solve combinatorial optimisation problems in manufacturing},
  author={Zhang, Cong and Wu, Yaoxin and Ma, Yining and Song, Wen and Le, Zhang and Cao, Zhiguang and Zhang, Jie},
  journal={IET Collaborative Intelligent Manufacturing},
  volume={5},
  number={1},
  pages={e12072},
  year={2023},
  publisher={Wiley Online Library},
 doi={10.1049/cim2.12072},
url={https://doi.org/10.1049/cim2.12072}

}

@misc{pennylane,
      title={PennyLane: Automatic differentiation of hybrid quantum-classical computations}, 
      author={Ville Bergholm and Josh Izaac and Maria Schuld and Christian Gogolin and Shahnawaz Ahmed and Vishnu Ajith and M. Sohaib Alam and Guillermo Alonso-Linaje and B. AkashNarayanan and Ali Asadi and Juan Miguel Arrazola and Utkarsh Azad and Sam Banning and Carsten Blank and Thomas R Bromley and Benjamin A. Cordier and Jack Ceroni and Alain Delgado and Olivia Di Matteo and Amintor Dusko and Tanya Garg and Diego Guala and Anthony Hayes and Ryan Hill and Aroosa Ijaz and Theodor Isacsson and David Ittah and Soran Jahangiri and Prateek Jain and Edward Jiang and Ankit Khandelwal and Korbinian Kottmann and Robert A. Lang and Christina Lee and Thomas Loke and Angus Lowe and Keri McKiernan and Johannes Jakob Meyer and J. A. Montañez-Barrera and Romain Moyard and Zeyue Niu and Lee James O'Riordan and Steven Oud and Ashish Panigrahi and Chae-Yeun Park and Daniel Polatajko and Nicolás Quesada and Chase Roberts and Nahum Sá and Isidor Schoch and Borun Shi and Shuli Shu and Sukin Sim and Arshpreet Singh and Ingrid Strandberg and Jay Soni and Antal Száva and Slimane Thabet and Rodrigo A. Vargas-Hernández and Trevor Vincent and Nicola Vitucci and Maurice Weber and David Wierichs and Roeland Wiersema and Moritz Willmann and Vincent Wong and Shaoming Zhang and Nathan Killoran},
      year={2022},
      eprint={1811.04968},
      archivePrefix={arXiv},
      primaryClass={quant-ph},
      url={https://arxiv.org/abs/1811.04968}, 
}

@software{deepmind2020jax,
  title = {The {D}eep{M}ind {JAX} {E}cosystem},
  author = {DeepMind and Babuschkin, Igor and Baumli, Kate and Bell, Alison and Bhupatiraju, Surya and Bruce, Jake and Buchlovsky, Peter and Budden, David and Cai, Trevor and Clark, Aidan and Danihelka, Ivo and Dedieu, Antoine and Fantacci, Claudio and Godwin, Jonathan and Jones, Chris and Hemsley, Ross and Hennigan, Tom and Hessel, Matteo and Hou, Shaobo and Kapturowski, Steven and Keck, Thomas and Kemaev, Iurii and King, Michael and Kunesch, Markus and Martens, Lena and Merzic, Hamza and Mikulik, Vladimir and Norman, Tamara and Papamakarios, George and Quan, John and Ring, Roman and Ruiz, Francisco and Sanchez, Alvaro and Sartran, Laurent and Schneider, Rosalia and Sezener, Eren and Spencer, Stephen and Srinivasan, Srivatsan and Stanojevi\'{c}, Milo\v{s} and Stokowiec, Wojciech and Wang, Luyu and Zhou, Guangyao and Viola, Fabio},
  url = {http://github.com/google-deepmind},
  year = {2020},
}

@article{lyngfelt_2025,
  title = {Symmetry-informed transferability of optimal parameters in the quantum approximate optimization algorithm},
  author = {Lyngfelt, Isak and Garc\'{\i}a-\'Alvarez, Laura},
  journal = {Phys. Rev. A},
  volume = {111},
  issue = {2},
  pages = {022418},
  numpages = {15},
  year = {2025},
  month = {Feb},
  publisher = {American Physical Society},
  doi = {10.1103/PhysRevA.111.022418},
  url = {https://link.aps.org/doi/10.1103/PhysRevA.111.022418}
}

\end{document}